\documentclass{PoS}

\title{The path from finite to infinite volume:\\
Hadronic observables from lattice QCD}

\ShortTitle{The path from finite to infinite volume}

\author{\speaker{Zohreh Davoudi}\thanks{Wilson award session.}\\
        Maryland Center for Fundamental Physics and Department of Physics,\\
University of Maryland, College Park, MD 20742, USA, and\\
RIKEN Center for Accelerator-based Sciences,
Wako 351-0198, Japan\\
        E-mail: \email{davoudi@umd.edu}}

\abstract{Standard Model determinations of properties of strongly interacting systems of hadrons have become possible with the powerful method of lattice quantum chromodynamics (LQCD), a method with  growing applicability and reliability. While growth in computational power and innovations in algorithmic and computational approaches have been essential in advancing the state of the field, conceptual and formal developments have played a crucial role in turning the output of LQCD computations to phenomenologically valuable results. From the invention of finite-volume technology to access physical observables by Martin L\"uscher over three decades ago to date, this field has grown in scope and complexity, enabling studies of scattering amplitudes and reaction rates, as well as spectroscopy of excited states of quantum chromodynamics (QCD) and resonances. Further, LQCD studies are augmented with the inclusion of quantum electrodynamics (QED), and subtleties related to the finite volume of systems in presence of QED have been understood and largely controlled. In this talk, I focus on selected developments to give a taste of the status of the field concerning the mapping between the finite and infinite-volume physics and its state-of-the-art applications.}

\FullConference{The 36th Annual International Symposium on Lattice Field Theory - LATTICE2018\\
		22-28 July, 2018\\
		Michigan State University, East Lansing, Michigan, USA.}

\begin{document}

%%%%%%%%%%%%%%%%%%%%%%%%%%%%%%%%%%%%%%%%%%%%%%%%%%%%%%%%%%%%%%%%%%%%%%%%%%%%%%%%%%%%%%%%%%%%%%%%%%%%%%
\section{Introduction}
Lattice quantum chromodynamics (LQCD) has become a powerful tool in first-principles studies of hadronic and nuclear phenomena. It has enabled reliable determinations based on the Standard Model (SM) of particle physics that have confronted experiment, or aim to open a window to  mechanisms beyond the SM. Among the accomplishments of this field are precise determinations of the lowest-lying hadron spectra including small mass splittings among members of hadron multiplets, the strong fine structure constant and quark masses, insights into the QCD phase diagram, hadronic contributions to the weak decay of mesons to constrain the elements of the CKM matrix, CP violation parameters in kaon systems, hadronic contributions to muon anomalous magnetic moment, hadron structure, hadron scattering and decays, resonance spectroscopy and structure, nuclear spectroscopy, structure, scattering and reactions, albeit yet at unphysically large quark masses when it comes to multi-baryon systems, see Refs.~\cite{Aoki:2016frl, Marinkovic:2017yzl, Ratti:2018ksb, Nocera:2017war, Briceno:2017max, Dudek:2018wai, Green:2018vxw, Davoudi:2017ddj} and other recent reviews. These accomplishments, aside from impressive advances in computational technologies and algorithms, reflect on the advanced state of the field with regard to conceptual and formal developments, without which the connection between numerical output of LQCD and physical observables, particularly in multi-hadron sector, would have been obscure.

In LQCD, the vacuum expectation values of observables of interest are evaluated in the background of the strongly interacting vacuum of quantum chromodynamics (QCD). LQCD relies on evaluating a quantum-mechanical path integral on a discretized and finite spacetime volume. The \emph{paths} or namely configurations are selected through a Monte Carlo importance sampling procedure, made possible by performing a Wick rotation to Euclidean spacetime. This gives rise to a thermodynamical interpretation of the path integral and allows the expectation values to be estimated from a statistical averaging over ensembles of vacuum configurations. The downside of this approach is that real-time dynamical observables cannot be simply constructed from an analytic continuation of the infinite-volume limit of LQCD correlation functions~\cite{Maiani:1990ca}. In the case of elastic $2 \to 2$ scattering, a method developed by L\"uscher circumvented this problem by providing a mapping between the finite-volume (FV) spectra of two particles and their scattering amplitude in infinite volume~\cite{Luscher:1990ux, Luscher:1986pf}. The intuitive picture is that the same finite-range interactions that constrain the scattering amplitude of two hadrons in QCD, give rise to a shift in energies of two interacting hadrons in a finite volume compared with free hadrons. So as long as the range of interactions is small compared with the extent of the volume, these two quantities must be ultimately related. Finding such relation in general requires a case-by-case study, hence efforts have been made to generalize L\"uscher's formula to more than two particles and to decay and reaction processes, as will be discussed in this proceedings.

Understanding the relation between finite and infinite-volume observables in the two-nucleon sector has led to ways to enhance the sensitivity of quantities that can be accessed through LQCD calculations such as energy spectra, to small but physically significant quantities such as the s-d mixing parameter in spin-triplet two-nucleon scattering. Further, analytical results on the volume dependence of observables such as binding energies have allowed for the development of improvement schemes, such as selected boundary conditions (BCs), to minimize the FV corrections incurred in a given LQCD simulation. This proceedings will include few examples of the benefits of a FV formalism beyond what is offered by L\"uscher's formula.  

As quantum electrodynamics (QED) plays an important role in certain hadronic quantities, in particular when electrically charged states are concerned, it is necessary to include QED interactions in LQCD studies. Since QED affects charged quarks, both in valence and sea sectors of QCD, only a nonperturbative inclusion of QED can provide reliable estimates for QED-sensitive observables, such as the mass splitting between hadron multiplets and charged-particle scattering. Nonetheless, for certain quantities, such as the response of hadronic states to external electromagnetic fields (EM), it may suffice to introduce a classical $U(1)$ field. Multitude of results have been obtained for magnetic moment and polarizabilities of hadrons and nuclei using such a method. In this proceedings, a framework will be presented to extend the background-field method to access quantities beyond what is considered to date, using the connection between FV and infinite-volume correlation functions in presence of general background fields. Additionally, frameworks for fully dynamical inclusion of $U(1)$ gauge fields in a finite volume will be presented. Although subtleties arise in embedding charged states in a finite volume given the infinite-range of QED interactions, remedies exist to still enable reliable LQCD+LQED studies of hadrons, one of which will be discussed in more detail in this proceedings.

%%%%%%%%%%%%%%%%%%%%%%%%%%%%%%%%%%%%%%%%%%%%%%%%%%%%%%%%%%%%%%%%%%%%%%%%%%%%%%%%%%%%%%%%%%%%%%%%%%%%%%
\section{Two-body elastic scattering}
%%%%%%%%%
%
\begin{figure}[t!]
\begin{center}  
\includegraphics[scale=0.475]{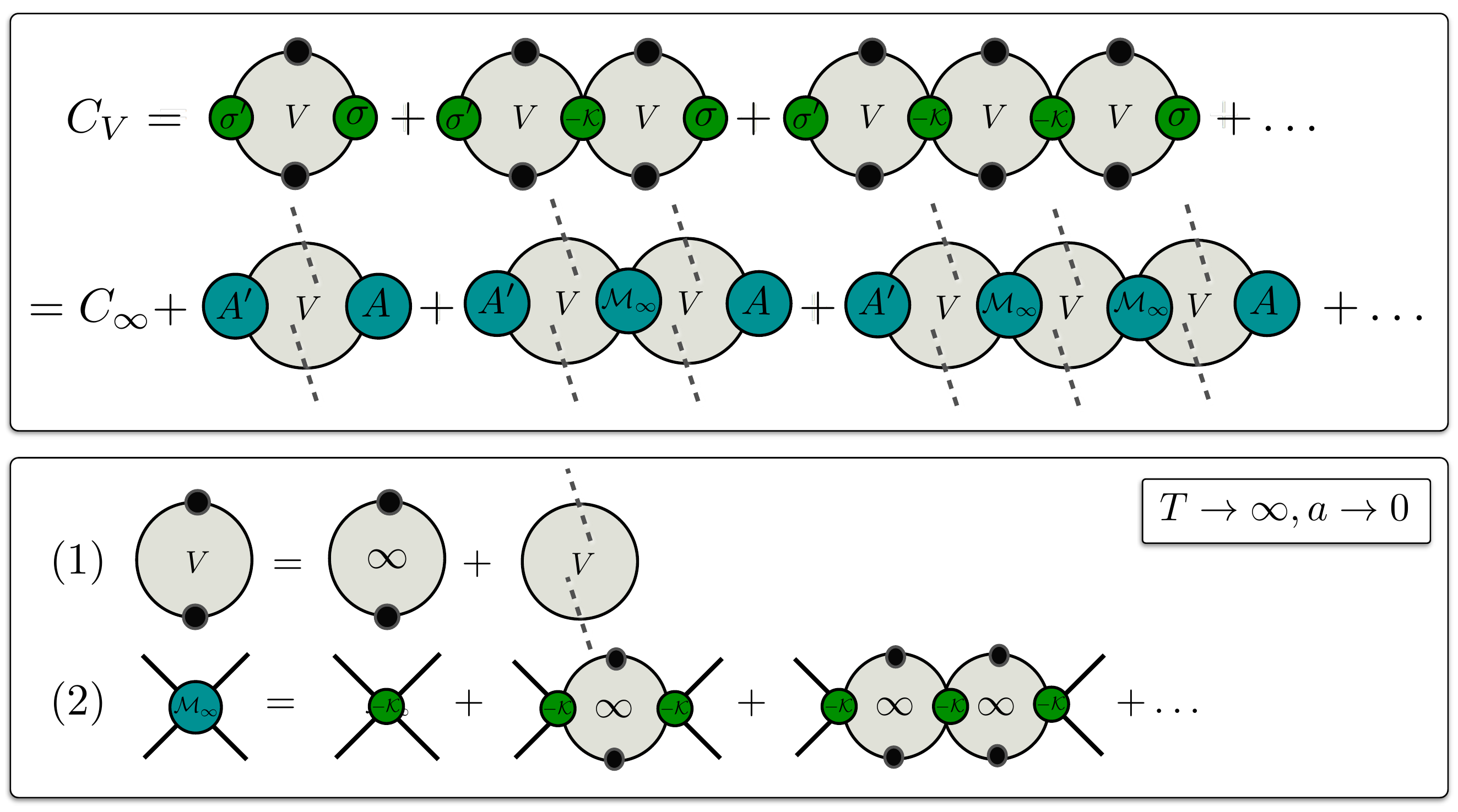}
\caption[.]{Diagrammatic expansion of $2 \to 2$ correlation function in a finite volume. $\sigma$ ($\sigma'$) is the interpolating operator for annihilation (creation) of a two-hadron state in a finite volume. $A$ ($A'$) is the physical transition amplitude between the vacuum and the two-hadron state induced by operator $\sigma$ ($\sigma'$). The lines are free propagators while the lines with a black dot on them denote fully dressed propagators. The $V$ letter inside the loops indicates that a summation over discrete momentum is assumed. Such loops can be replaced by their infinite-volume counterparts as well as a purely FV contribution, up to exponential corrections in volume. The FV contribution is specified by loops in which the propagators are crossed with dashed lines, indicating that they are evaluated at on-shell kinematics, giving rise to power-law corrections in volume. $\mathcal{M}_{\infty}$ denotes the physical on-shell $2 \to 2$ scattering amplitude, displayed in the lower panel as an infinite sum over  $2 \to 2$ s-channel diagrams with the interacting kernel, $\mathcal{K}$. The kernel, as well as the fully-dressed propagators, can be replaced with their infinite-volume counterparts below the inelastic threshold, up to exponential corrections in volume.}
\label{fig:cv}
\end{center}
\end{figure}
\noindent
The two-particle scattering amplitude has a wealth of information about the spectrum of a theory. It exhibits poles at real values of energy if interactions permit two-particle bound states, or at complex values of energy if there are resonances. Further, a cut starts at the scattering threshold. Such a rich content needs to arise from a LQCD calculation of two-hadron systems in a finite Euclidean spacetime that provides only a discrete set of lowest-lying energy eigenvalues.

To find the relation between the scattering amplitude and the FV spectrum, let us start from the diagrammatic expansion of the correlation function shown in Fig.~\ref{fig:cv}. The infinite sum over all s-channel two-hadron loops connected via the s-channel $2 \to 2$ Bethe-Salpeter kernels in a finite volume can be rewritten as the infinite-volume part, plus a summation over two-hadron loops evaluated on-shell and connected via the physical on-shell $2 \to 2$ scattering amplitude, up to corrections that scale as $e^{-RL}$, where $L$ denotes the spatial extent of the cubic volume and $R$ is the finite range of interactions (the pion Campton wavelength in QCD)~\cite{Luscher:1985dn, Bedaque:2006yi}. The on-shell condition decouples the loops in the latter summation, up to an angular dependence that can be projected to partial waves, see Refs.~\cite{Kim:2005gf, Briceno:2013lba}. This will give rise to a geometric series of matrices in angular momentum basis that can be summed up to all orders. Noting that the poles of the FV correlation function correspond to energy eigenvalues of two hadrons in a finite volume leads to the FV quantization condition (QC) or L\"uscher's formula,
\begin{eqnarray}
\det(\mathcal{M}^{-1}+\delta \mathcal{G}^{V})=0,
\label{eq:QC}
\end{eqnarray} 
where 
\begin{eqnarray}
(\delta \mathcal{G}^{V})_{l_1,m_1;l_2,m_2}&=& i\frac{q^*n}{8\pi E^*}\left(\delta_{l_1,l_2}\delta_{m_1,m_2}+i\frac{4\pi}{q^*}\sum_{l,m}\frac{\sqrt{4\pi}}{q^{*l}}c^{\textbf{P}}_{lm}(q^{*2})\int d\Omega Y^*_{l_1m_1}Y^*_{lm}Y_{l_2m_2}\right).
\end{eqnarray}
$n$ is ${1}/{2}$ if the two hadrons are identical and is 1 otherwise. $\mathbf{P}$ is the total three-momentum of the system, $E^*$ is the total center of mass (CM) energy and $q^*$ is the momentum of each hadron in the CM frame on shell. The function $c^{\textbf{P}}_{lm}$ is defined as
\begin{eqnarray}
c^{\textbf{P}}_{lm}(x)=\frac{1}{\gamma}\left[\frac{1}{ L^3}\sum_{\textbf{k}}-\mathcal{P}\int\frac{d^3\mathbf{k}}{(2\pi)^3}\right]\frac{\sqrt{4\pi}Y_{lm}(\hat{k}^*)~k^{*l}}{{k}^{*2}-x}.
\label{eq:clm}
\end{eqnarray}
$\mathcal{P}$ in this relation denotes the principal value of the integral, and $\mathbf{k}^*={\gamma}^{-1}(\mathbf k_{||}-\alpha \mathbf P)+\mathbf k_{\perp}$, where $\mathbf k_{||}$ ($\mathbf k_{\perp}$) denotes the component of the momentum vector $\mathbf{k}$ that is parallel (perpendicular) to the boost vector $\mathbf{P}$. $\gamma$ is the relativistic boost factor and $\alpha=\frac{1}{2}\left[1+\frac{m_1^2-m_2^2}{E^{*2}}\right]$, where $m_1$ and $m_2$ are the masses of the hadrons~\cite{Davoudi:2011md, Fu:2011xz, Leskovec:2012gb}. This result is equivalent to the result obtained in Refs. \cite{Rummukainen:1995vs, Kim:2005gf, Christ:2005gi} for boosted systems with identical masses.
\begin{figure}[h!]
\begin{center}  
\includegraphics[scale=0.555]{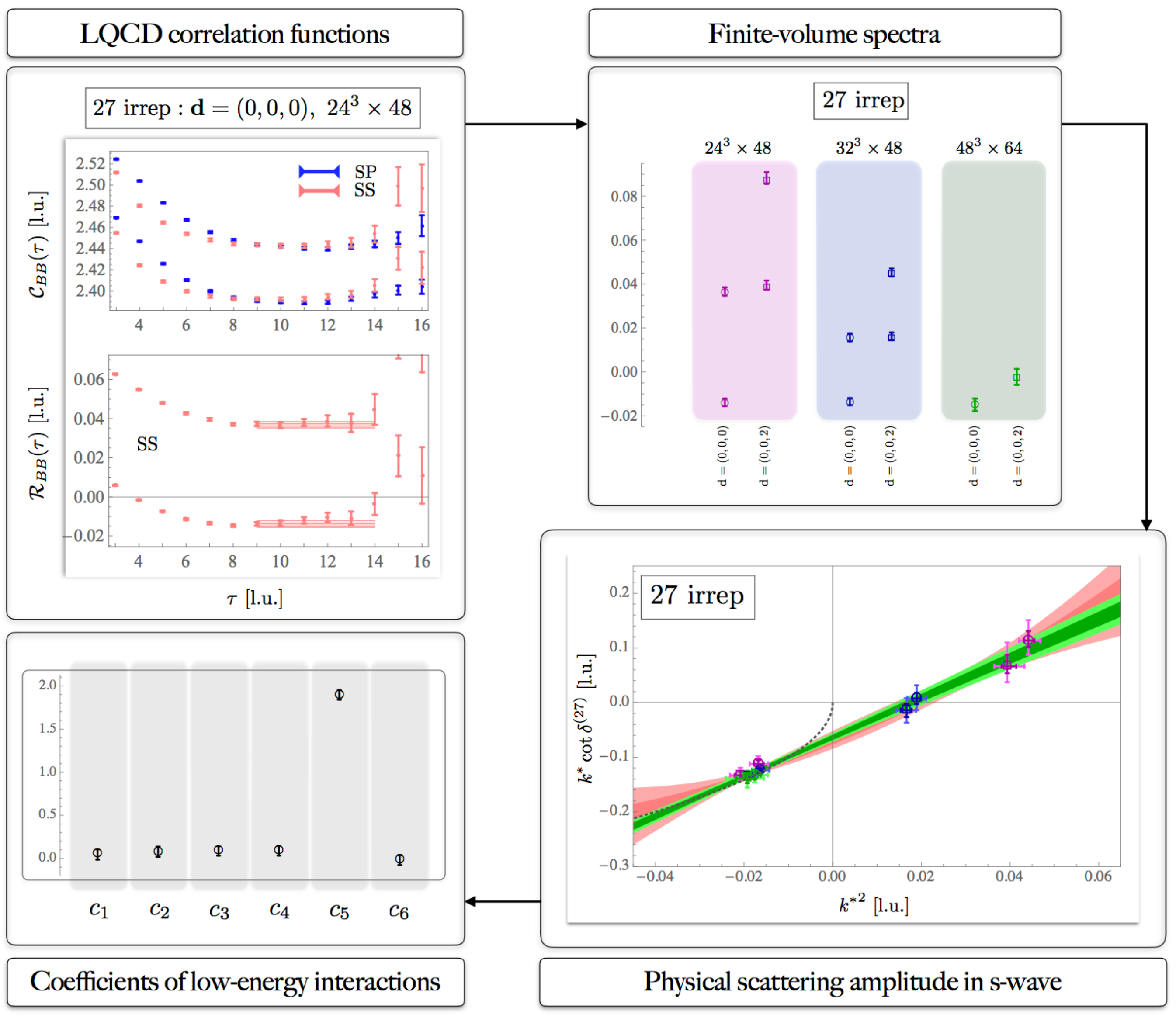}
\caption[.]{Baryon-baryon scattering from LQCD at a $SU(3)$ flavor-symmetric point with $m_{\pi} \approx 800~\tt{MeV}$~\cite{Wagman:2017tmp}. The figure in the upper-left panel exhibits a function of correlators in the $27$ irreducible representation (irrep) of $SU(3)$ to which the $NN~{^1}S_0$ channel belongs. From the late-time behavior of these correlators, the lowest-lying spectra of the system were obtained at several lattice volumes and total CM boosts, as shown in the upper-right panel of the figure (SS and SP denote two different sink operators used). L\"uscher's FV QC turned this spectra to constraints on the s-wave scattering phase shifts at the corresponding energies, which were fit to the EREs with two (green band) and three (pink band) parameters, as depicted in the lower-right panel. Finally, the constraints on the scattering amplitude in this channel, and five additional channels, were turned into constraints on the leading $SU(3)$ low-energy constants (LECs) of the EFT (shown in the lower-left panel), most strikingly pointing to an emergent $SU(16)$ extended symmetry in two octet baryon interactions, predicted to exist in the large-$N_c$ limit of QCD~\cite{Kaplan:1995yg}, and observed for the first time in Ref~\cite{Wagman:2017tmp}.}
\label{fig:bb800}
\end{center}
\end{figure}

The QC in Eq.~(\ref{eq:QC}) is an infinite-dimensional nondiagonal matrix due to the breakdown of rotational symmetry in a cubic volume. At low energies, the condition can be truncated to a smaller matrix and parameterizations of the amplitude as a function of energy can be constrained. Fig.~\ref{fig:bb800} demonstrates a recent application of L\"uscher's QC that led to constraints on the s-wave scattering phase shifts in various two octet-baryon channels, albeit at an unphysically large value of the quark masses. This study, which is an example of the matching between LQCD determination of scattering observables and the corresponding effective field theory (EFT) description of interactions, pointed to a strong hint of spin-flavor symmetry in nuclear and hypernuclear interactions, as well as an extended accidental symmetry, that were both predicted in Ref.~\cite{Kaplan:1995yg} to arise from the large-$N_c$ limit of QCD. See also Ref.~\cite{Beane:2018oxh} for an interesting recent explanation.

%%%%%%%%%%%%%%%%%%%%%%%%%%%%%%%%%%%%%%%%%%%%%%%%%%%%%%%%%%%%%%%%%%%%%%%%%%%%%%%%%%%%%%%%%%%%%%%%%%%%%%
\section{Two-body inelastic scattering}
\begin{figure}[t!]
\begin{center}  
\includegraphics[scale=0.575]{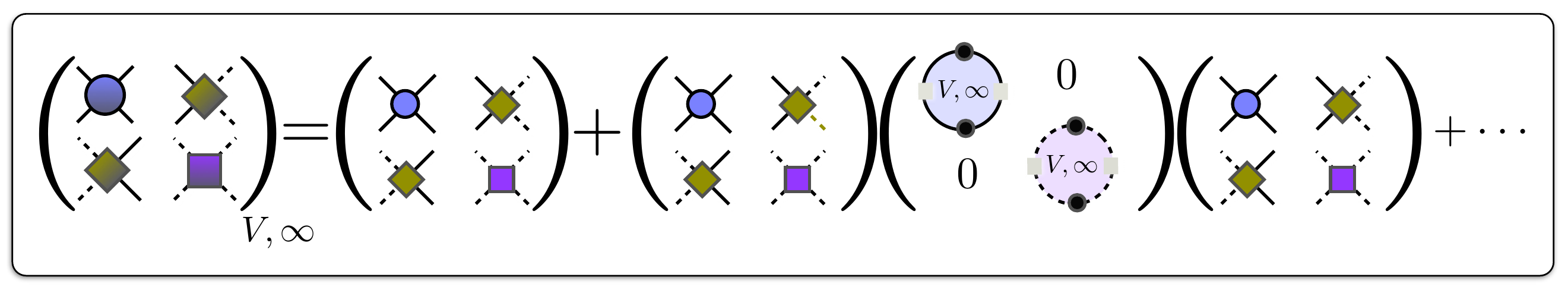}
\caption[.]{Diagrammatic expansion of the coupled-channels correlation function~\cite{Briceno:2012yi}. The solid (dashed) line is the propagator of each hadron in channel I (II), and black dot on the lines indicates that the propagator is fully dressed. The blue circle (purple square) is the interacting $2 \to 2$ kernel that sums up all s-channel two-particle irreducible diagrams for channel I (II), and the green diamond denotes the interacting $2 \to 2$ s-channel kernel coupling channel I and II. The $V$ letter ($\infty$ symbol) inside the loops implies that a sum (integral) over discrete (continuous) momentum is performed.}
\label{fig:cc}
\end{center}
\end{figure}
\noindent
For $2 \to 2$ processes in which multiple two-hadron channels are involved, not only does the QC need to obtain the elastic scattering amplitude in each channel, but also it should give access to the off-diagonal elements of the scattering matrix, i.e., those characterizing transitions among different two-body channels when there is sufficient energy available. An example of such coupled channels is the $\pi\pi-K\bar{K}-\eta\eta$ system, in which phenomenologically interesting resonances such as $\sigma$ and $f_0$ occur. In fact, a recent state-of-the-art LQCD study~\cite{Briceno:2017qmb} has implemented the generalization of the L\"uscher's QC to the case of multiple coupled channels~\cite{Briceno:2012yi, Hansen:2012tf} to obtain scattering parameters in this system (at unphysically larger values of the quark masses). This generalized formula can be obtained most straightforwardly from promoting the correlation function in Fig.~\ref{fig:cv} to a matrix, as shown in Fig.~\ref{fig:cc}. It is then easy to see that the QC remains in the form of Eq.~(\ref{eq:QC}), except for the fact that the scattering amplitude is now a matrix in the space of flavor channels with nonzero off-diagonal elements. Similarly, the FV matrix $\delta \mathcal{G}$ is a matrix in such space but is in diagonal form. The coupled-channels system does not need to be consisting of different flavor states, it may as well be a coupled system in the angular momentum basis, as is the case in two-nucleon systems in spin-triplet configurations, see Sec.~\ref{sec:NN}.

\begin{figure}[t!]
\begin{center}  
\includegraphics[scale=0.575]{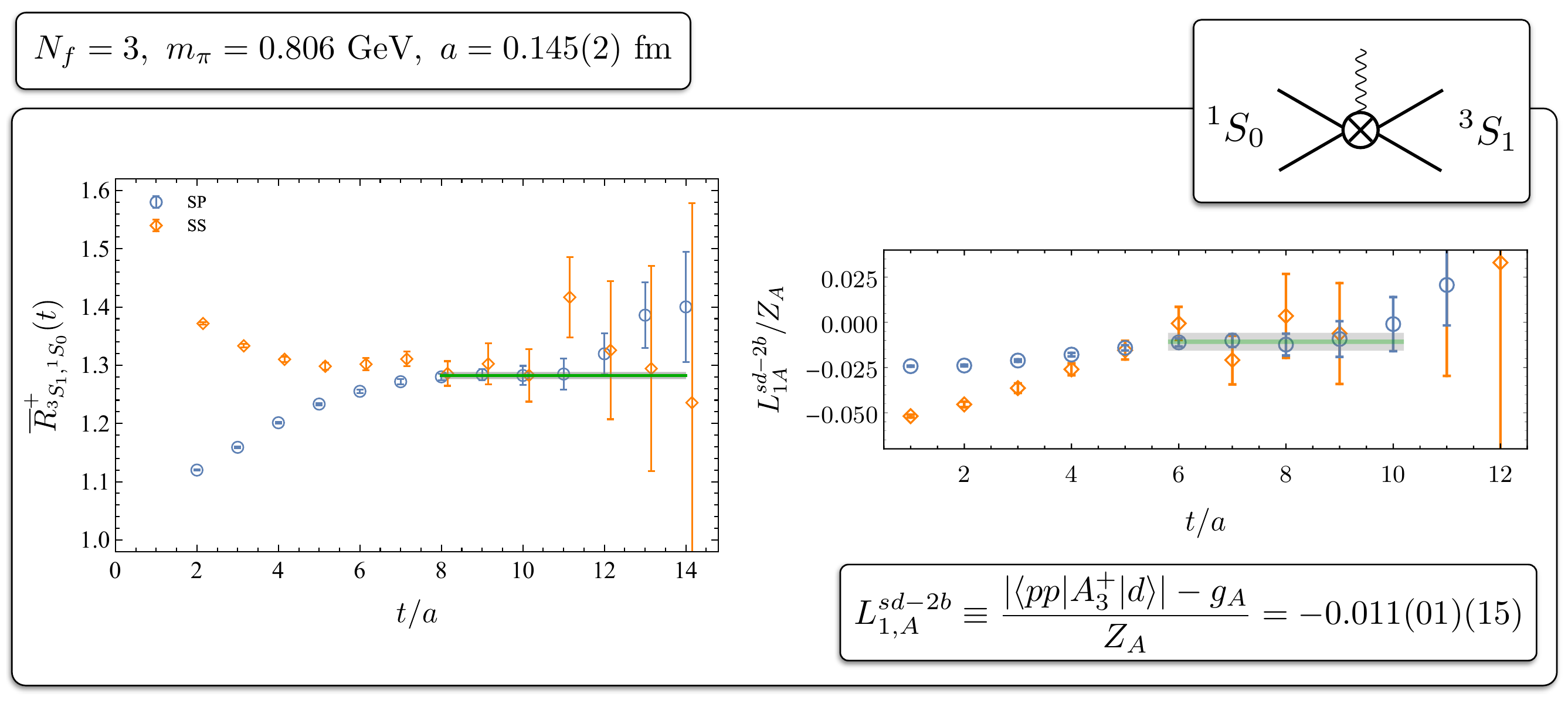}
\caption[.]{A LQCD calculation at a $SU(3)$ flavor-symmetric point with $m_{\pi} \approx 800~\tt{MeV}$ obtains the matrix element for the axial vector transition $NN~{^1}S_0 \to NN~{^3}S_1$ (left panel), leading to a constraint on the short-distance solely two-nucleon contribution to the matrix element, $L_{1,A}^{sd-2b}$, and the two-nucleon axial-vector coupling of the pionless EFT, $L_{1,A}$, at the physical values of the quark masses, assuming a mild dependence of the former on the quark masses. For further details, see Refs.~\cite{Savage:2016kon, Tiburzi:2017iux}. The formalism discussed in this section will be essential in connecting this matrix element to the physical amplitude at lower quark masses when the initial/final states are only nearly bound or unbound.}
\label{fig:pp}
\end{center}
\end{figure}
An interesting application of the coupled-channels FV formalism is the derivation of a Lellouch-L\"uscher formula for $2 \to 2$ amplitudes~\cite{Lellouch:2000pv}. The original Lellouch-L\"uscher formula provides a mapping between the physical decay amplitude of a $1 \to 2$ process and the corresponding matrix element in a LQCD calculation. It later enabled precision QCD determinations of the decay width of $K \to \pi \pi$~\cite{Boyle:2012ys, Blum:2015ywa}. The formula can be written as
\begin{eqnarray}
\frac{\left|_{\infty}\big\langle \pi(E_n/2,\vec{q}_n),\pi(E_n/2,-\vec{q}_n)| H_w |K(m_K,\vec{0})\big \rangle_{\infty}\right|^2}{\left|_{V}\big\langle \pi(E_n/2,\vec{q}_n),\pi(E_n/2,-\vec{q}_n)| H_w |K(m_K,\vec{0})\big \rangle_{V}\right|^2}
=\frac{32\pi^2L^3}{ E_n|\vec{q}_n|}(\delta'+\phi'),
\label{eq:LL}
\end{eqnarray}
in a frame where kaon is at rest. Here, $E_n$ is a FV eigenenergy of the final-state pions such that $m_K=E_n$. Ways to achieve this is to tune the volume or implement certain BCs. $q_n$ is the corresponding on-shell momentum of each pion. $\delta$ is the two-pion scattering phase shift in s wave and $\phi$ is a kinematic function related to $c^{\textbf{P}}_{lm}(x)$ in Eq.~(\ref{eq:clm}). Prime denotes derivative with respect to the energy. For conventions regarding the normalization of states and the interaction Hamiltonian $H_w$, see Ref.~\cite{Christ:2015pwa}.

Another phenomenologically interesting process in nuclear physics is proton-proton ($pp$) fusion, whose rate is hard to measure in experiment at low incident velocities which is relevant to energy production in Sun. One of the goals of the LQCD research in nuclear physics is to constrain the rate of this process and other reaction processes in the few-nucleon sector. As a step towards this gaol, Ref.~\cite{Savage:2016kon} combines a pionless EFT~\cite{Chen:1999tn} description of the threshold amplitude for $pp$ fusion~\cite{Butler:2001jj} and the LQCD determination of the matrix element of the axial-vector current between isosinglet and isotriplet two-nucleon states at an unphysical value of the quark masses to constrain the short-distance two-nucleon axial-vector coupling of the EFT, the so-called $L_{1,A}$. Given the large values of the quark masses, the initial and final two-nucleon states in this study were rather deeply bound. As a result, the FV matrix element could be taken to be the physical infinite-volume value up to small corrections that scale $e^{-\kappa L}$, where $\kappa$ is the binding momentum of the initial/final state. Towards the physical point, this feature is lost and the use of a generalized Lellouch-L\"uscher formula to access the physical amplitude will be required.

This generalized formula, as stated above, can be derived from the QC for the coupled two-nucleon systems. A further complication is the presence of single-body contributions to the process, i.e., a proton converting to a neutron, which nonetheless can be accounted for in the formalism straightforwardly, see Ref.~\cite{Briceno:2012yi}. The resulting relation between the FV matrix elements of the Hamiltonian density, $\mathcal{M}^V_{W}$, and the two-nucleon scale-independent LECs of the pionless EFT, $\widetilde{L}_{1,A}$, at leading order (LO) in the weak interaction is
\begin{eqnarray}
\left(\frac{mV}{2}\right)^2\csc^2{\delta^{({^1}S_0)}}\csc^2{\delta^{({^3}S_1)}}
\left(\phi'+\delta^{({^1}S_0)'} \right)\left(\phi'+\delta^{({^3}S_1)'}\right)|\mathcal{M}^V_{W}|^2
=\left(\frac{4\pi}{m}
{\widetilde{L}_{1,A}}+\frac{4\pi}{m}g_AG_1^V\right)^2.
\label{eq:LLNN}
\end{eqnarray}
Here, $G_1^V$ in related to the derivative of $c^{\textbf{P}}_{lm}(x)$ with respect to $x$, $\delta^{({^1}S_0)}$ ($\delta^{({^1}S_0)}$) is the s-wave scattering phase shift in the $NN~{^1}S_0$ ($NN~{^3}S_1$) channel, $V=L^3$ is the spatial volume, $m$ is the mass of each nucleon and $g_A$ is the nucleon's axial charge. More general Lellouch-L\"uscher formulae that make no reference to the underlying EFT and assume generic one-body and two-body currents, were developed later in Refs.~\cite{Briceno:2015tza, Briceno:2015csa, Baroni:2018iau}.

%%%%%%%%%%%%%%%%%%%%%%%%%%%%%%%%%%%%%%%%%%%%%%%%%%%%%%%%%%%%%%%%%%%%%%%%%%%%%%%%%%%%%%%%%%%%%%%%%%%%%%
\section{Two-nucleon observables
\label{sec:NN}}
\begin{figure}[t!]
\begin{center}  
\includegraphics[scale=0.655]{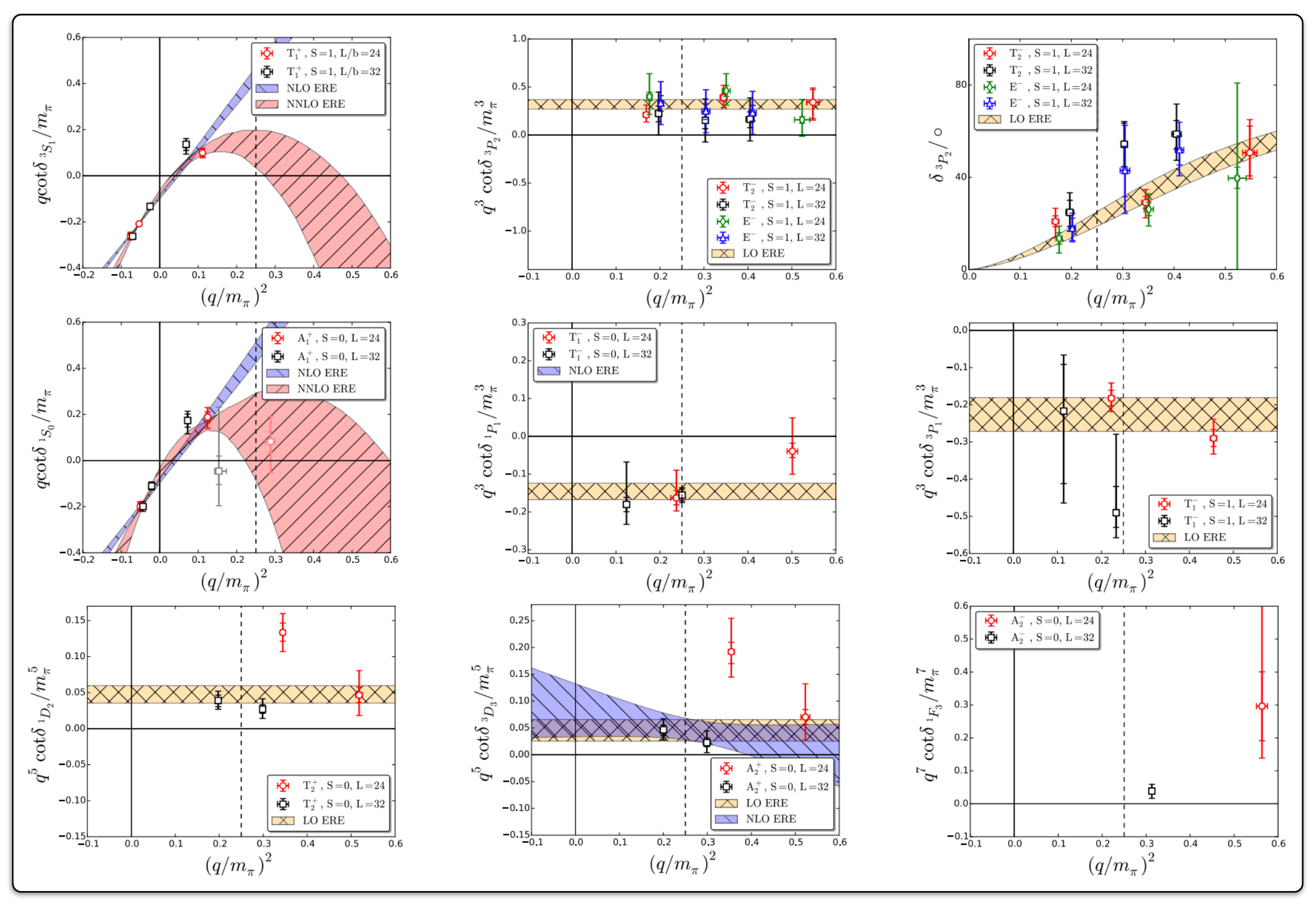}
\caption[.]{QCs in two-nucleon systems~\cite{Briceno:2013lba} led to constraints on scattering phase shifts of two-nucleon systems in higher partial waves directly from QCD, albeit at large quark masses corresponding to $m_{\pi} \approx 800$ MeV~\cite{Berkowitz:2015eaa}. See Ref.~\cite{Berkowitz:2015eaa} for further detail.}
\label{fig:partialwaves}
\end{center}
\end{figure}

\noindent
There are far more information accessible in L\"uscher's QC than just the s-wave scattering phase shift. A particularly valuable development was the generalization of L\"uscher's QC to two-nucleon systems with given spin and isospin, and its decomposition to several independent QCs corresponding to given irreducible representations (irreps) of the $O$, $D_4$ and $D_2$ point groups, corresponding to two-nucleon systems at rest, and with boost vectors $d=(0,0,1)$ and $(1,1,0)$, respectively~\cite{Briceno:2013lba}. These QCs were later used in Ref.~\cite{Berkowitz:2015eaa} to obtain the scattering phase shifts of two-nucleon systems in higher partial waves, as depicted in Fig.~\ref{fig:partialwaves}.
\begin{figure}[t!]
\begin{center}  
\includegraphics[scale=0.545]{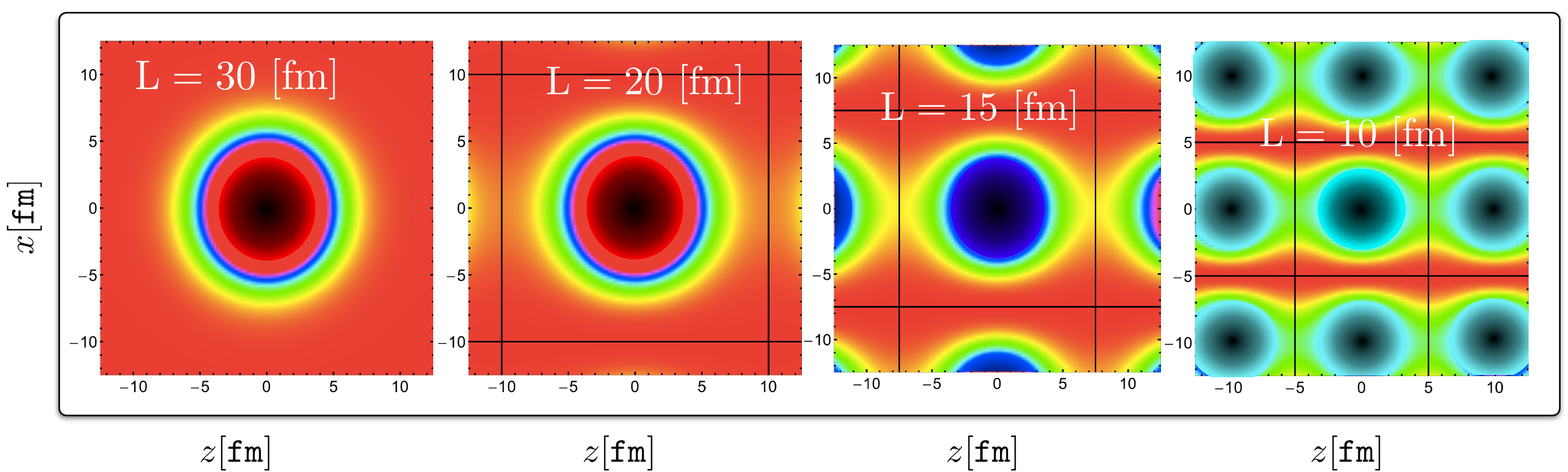}
\caption[.]{The mass density in the $x$-$z$ plane from the FV deuteron wavefunction with $d=(0,0,1)$ in the $\mathbb{E}$ irrep of the $D_4$ point group. Distortion of the tail of the wavefunction along the direction of the boost in a finite volume is evident.}
\label{fig:wavefunction}
\end{center}
\end{figure}
\begin{figure}[t!]
\begin{center}  
\includegraphics[scale=0.555]{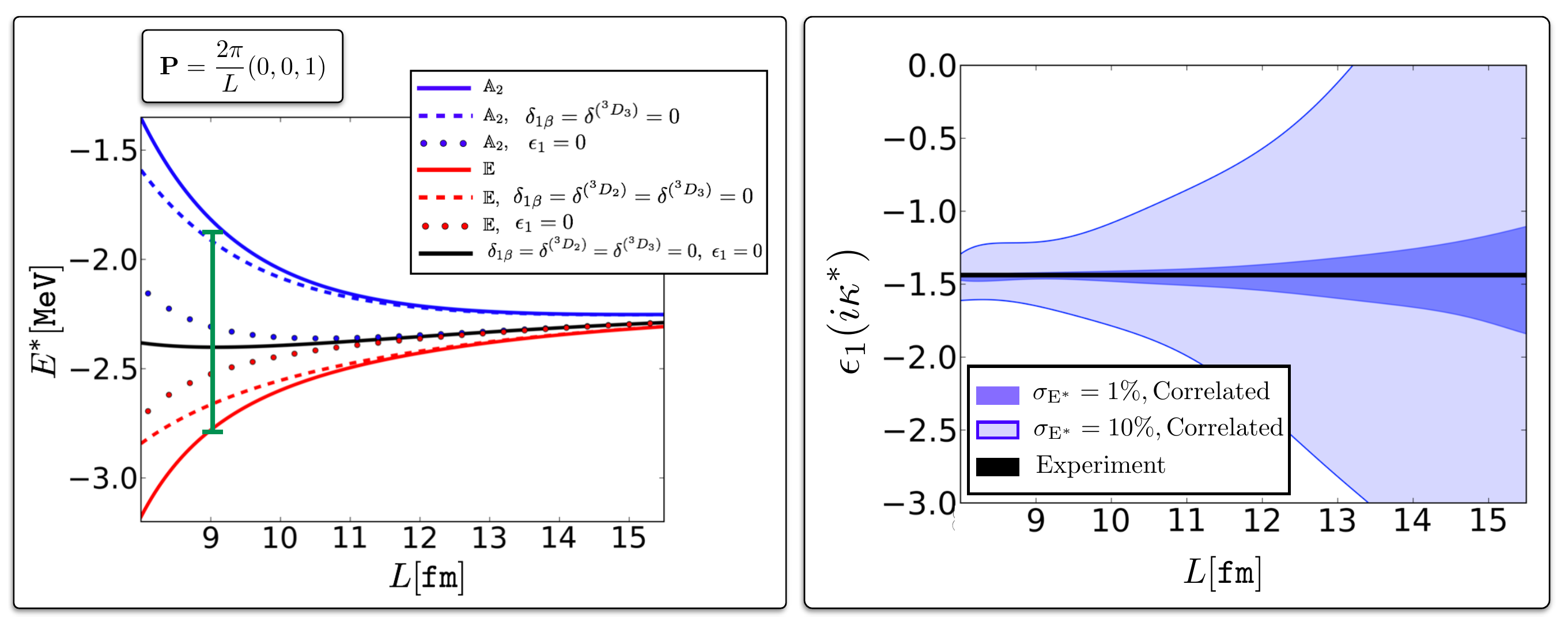}
\caption[.]{The left panel shows the ground-state energy of two nucleons in the $NN~{^3}S_1-{^3}D_1$ channel with $d=(0,0,1)$ as a function of $L$, in two irreps of the $D_4$ point group~\cite{Briceno:2013bda}. The difference between the two-nucleon energies in a finite volume, as is seen, is highly sensitive to the $\epsilon_1$ parameter characterizing the mixing between s-wave and d-wave partial waves in this channel. The right panel is the result of a fake data analysis, where a $1\%$ and $10\%$ precision is assumed on the FV energy levels and the relevant FV QC was used to extract the mixing parameter as a function of the spatial extent of the cubic volume, $L$. For this analysis, given low-energy parameterizations of the phase shifts and the mixing parameter had to be assumed, see Ref.~\cite{Briceno:2013bda} for detail.}
\label{fig:epsilon}
\end{center}
\end{figure}

Another interesting application of the two-nucleon QCs with given boost vectors is in constraining the small mixing parameter between partial waves in spin-triplet channels. An example is the two-nucleon channel that exhibits deuteron as a bound state. While predominantly s wave, the tensorial nuclear force induces a d-wave component in the deuteron wavefunction, making it a nonspherical state with a nonzero quadrupole moment, $Q_d = 0.2859(3) {\rm fm}^2$~\cite{PhysRevA.20.381}. The S-matrix for this coupled-channels system can be parameterized by two phase shifts, $\delta_{1\alpha}$ and $\delta_{1\beta}$, and one mixing angle, $\epsilon_1$~\cite{Blatt:1952zza,PhysRev.93.1387}, with the mixing angle manifesting itself in the asymptotic d/s ratio of the deuteron wavefunction, $\eta = 0.02713(6)$~\cite{PhysRev.93.1387, PhysRevC.47.473, deSwart:1995ui}. The question that arises is that given the current and upcoming achievable precision in the two-nucleon spectroscopy, can such a small d-wave component be identified in a LQCD study?

By inputting the experimentally known phase shifts and mixing parameters, the FV energy spectra of two-nucleon systems at rest is found rather insensitive to the small mixing parameter~\cite{Briceno:2013bda}. However, by boosting the nucleons in smaller volumes, e.g., along the $z$ axis (defined with respect to the spin direction of the deuteron), the deuteron wavefunction is further deformed, as seen in Fig.~\ref{fig:wavefunction}, and the mixing parameter induces a large splitting between the lowest energy eigenvalues in two different irreps of the corresponding point group, see the left panel of Fig.~\ref{fig:epsilon}. Using fake data, it has been verified that a percent-level determination of the mixing parameter is achievable with a percent-level extraction of the lowest-lying FV spectra~\cite{Briceno:2013bda}. It is also notable that once again, finite volume does not present a limitation but an advantage in accessing physical quantities of interest. As is seen in the right panel of Fig.~\ref{fig:epsilon}, on one hand the accuracy in the determination of $\epsilon_1$ may be lost at smaller volumes due to the onset of exponential volume corrections that are neglected in formalism, and on the other hand the precision is lost by tuning the volume to be too large.
\begin{figure}[b!]
\begin{center}  
\includegraphics[scale=0.575]{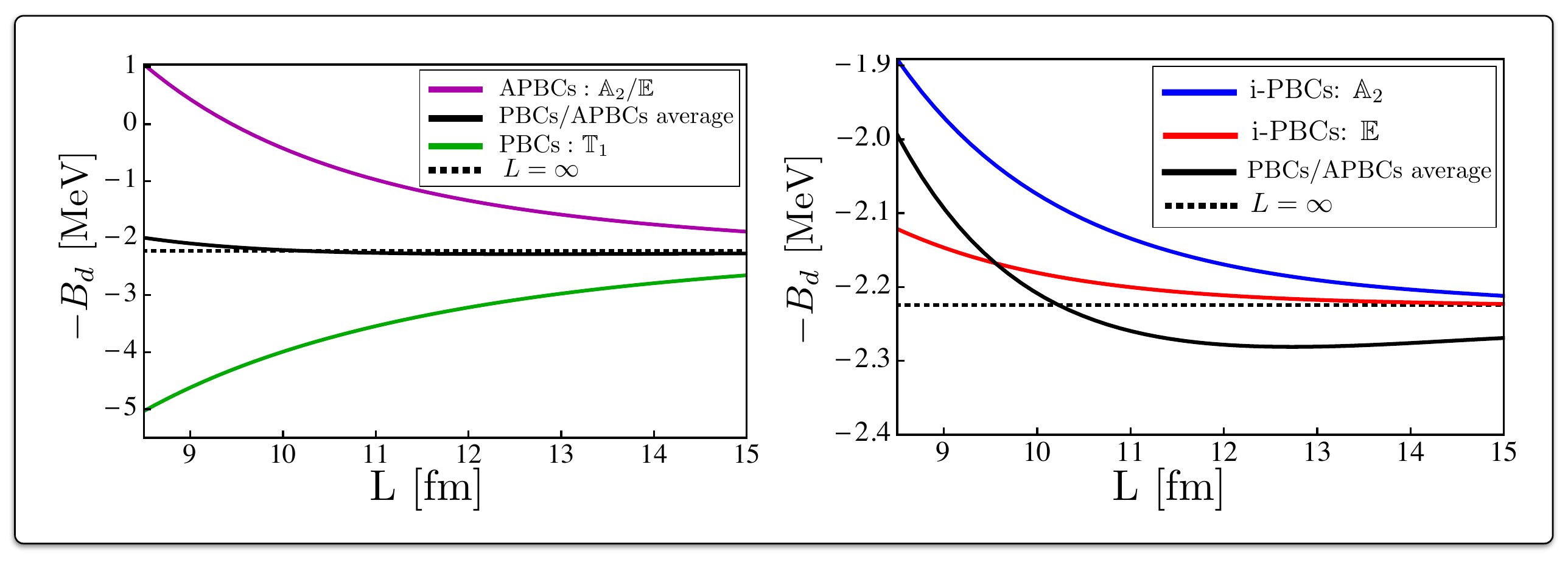}
\caption[.]{The deuteron binding energy as a function of $L$ evaluated with various BCs. In the left panel, periodic BCs (green curve), antiperiodic-BCs (purple curve) and their average (black curve) are compared against the infinite-volume deuteron binding energy (dashed line), clearly demonstrating the insensitivity of the average to the volume. In the right panel, this average is compared against the so-called i-periodic BCs, in which the wavefunction of the quarks is multiplied by $i$ after traversing the volume by one period in each spatial direction. As is seen, such a BC gives rise to very mild volume-dependence in the deuteron's energy even at moderately small volumes.}
\label{fig:tbc}
\end{center}
\end{figure}

Finally, while FV effects can be valuable in constraining some observable, there are situations where such effects should be considered as contamination and need to be removed before further analysis. One example is in studying bound states in a volume, where the tail of the wavefunction gets distorted as a result of boundary effects. While these effects are exponential in form~\cite{Bour:2011ef, Davoudi:2011md}, they can be large in a finite volume, e.g., the ground-state energy of two nucleons in the deuteron channel in a volume of $\sim (9~{\rm fm})^3$ is twice the physical value of the deuteron binding energy. While the FV formalism allows quantifying these effects, it can be advantageous to devise studies that introduce much less significant volume corrections in properties of bound states, for example where the scattering of a bound state with another hadron/bound state is considered. Twisted BCs~\cite{Bedaque:2004kc} introduce one such possibility for an improvement scheme~\cite{Briceno:2013hya}. As is shown in Fig.~\ref{fig:tbc}, encouraging results for the deuteron binding energy is observed for various choices of BCs on the quark fields.

%%%%%%%%%%%%%%%%%%%%%%%%%%%%%%%%%%%%%%%%%%%%%%%%%%%%%%%%%%%%%%%%%%%%%%%%%%%%%%%%%%%%%%%%%%%%%%%%%%%%%%
\section{Three-body elastic scattering}
\noindent
Constraining three-hadron scattering amplitudes directly from QCD is an overarching goal of the community, with motivations ranging from investigations of the breakdown of fundamental symmetries of nature in mesonic sector, excited-state spectroscopy and research on QCD resonances,   as well as three-neutron forces for research on neutron-rich isotopes and dense matter. Ideally, a model-independent procedure analogous to the two-body L\"uscher's method is desired to convert the spectral information in a finite volume to the three-body scattering amplitudes through a QC. Such a QC in fact exists in the literature~\cite{Polejaeva:2012ut,Briceno:2012rv,Hansen:2014eka,Hammer:2017uqm,Hammer:2017kms,Briceno:2017tce,Meng:2017jgx,Mai:2017bge,Briceno:2018aml}, although often in radically different forms, and its implementation appears more complex than the two-body QC~\cite{Kreuzer:2010ti,Mai:2018djl,Briceno:2018mlh}. An ongoing effort in the community is devoted to exploring the consistency of these forms and their various limits~\cite{Detmold:2008gh,Meissner:2014dea,Hansen:2016fzj,Konig:2017krd}, either formally, or through comparing their output for the FV spectra. Here, we briefly describe one of these approaches and refer the reader to other proceedings in this conference series for more recent work~\cite{Mai:2018xwa, Blanton:2018guq}.

\begin{figure}[t!]
\begin{center}  
\includegraphics[scale=0.635]{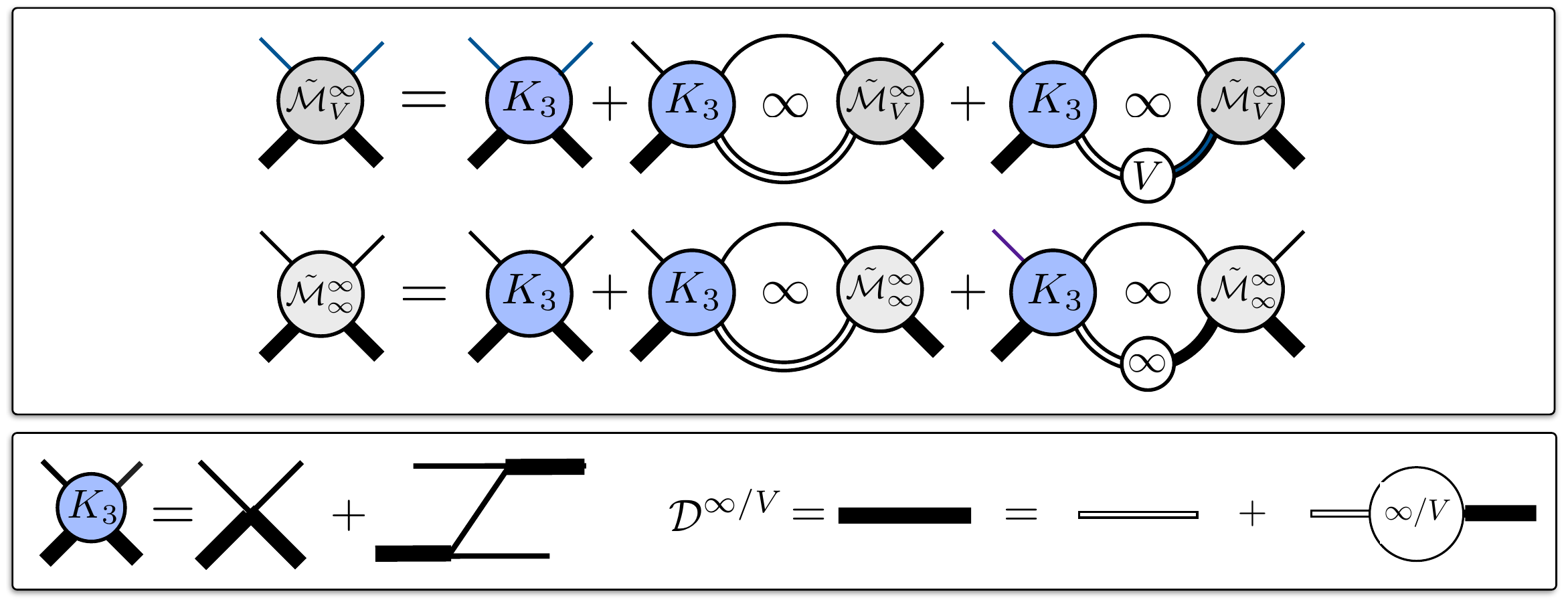}
\caption[.]{The QC in Eq.~(\ref{eq:3bQC-2}) contains the amplitudes shown in the upper panel. The ingredient of these amplitudes are the three-body interaction kernel at LO in the EFT (the blue blob), consisting of a contact three-body force and a particle-exchange diagram,  and the full dimer propagator (the thick black lines) arising from summing up all $2 \to 2$ interactions in the two-body sector, as shown in the lower panel. The thin black line is the nucleon propagator, and the black double-line is the free dimer propagator. The s-channel dimer-nucleon loops are evaluated either as a loop integral over momenta (corresponding to the $\infty$ symbol) or a sum over discrete spatial momenta in a finite volume (corresponding to the letter V).}
\label{fig:3b}
\end{center}
\end{figure}

Within a dimer-particle framework, a master relation derived in Ref.~\cite{Briceno:2012rv} to relate the three-body interactions at low energies (embedded in a three-body interaction kernel, $\mathcal{K}_3$, as shown in Fig.~\ref{fig:3b}) to the energy eigenvalues of three spinless hadrons in a finite volume. This can also be written in terms of a relation that involves the physical dimer-particle scattering amplitude in s-wave, $ \tilde{\mathcal{M}}^{\infty}_{\infty}$,
\begin{eqnarray}
&&\tilde{\mathcal{M}}^{\infty}_{V}\left(\mathbf{p},\mathbf{k};\mathbf{P},E\right) = K_3(\textbf{p},\textbf{k};{\mathbf{P}},E) -
\int\frac{d^3q}{(2\pi)^3}K_3(\textbf{p},\textbf{q};{\mathbf{P}},E){\mathcal{D}^{V}(E-\frac{q^2}{2m},|\mathbf{P}-\mathbf{q}|)} \tilde{\mathcal{M}}^{\infty}_{V}\left(\mathbf{q},\mathbf{k};\mathbf{P},E\right) \nonumber\\
&& \hspace{1.0 cm} =  
 \tilde{\mathcal{M}}^{\infty}_{\infty}(\textbf{p},\textbf{k};{\mathbf{P}},E) -
\int\frac{d^3q}{(2\pi)^3}\tilde{\mathcal{M}}^{\infty}_{\infty}\textbf{p},\textbf{q};{\mathbf{P}},E)\delta{\mathcal{D}^{V}(E-\frac{q^2}{2m},|\mathbf{P}-\mathbf{q}|)} \tilde{\mathcal{M}}^{\infty}_{V}\left(\mathbf{q},\mathbf{k};\mathbf{P},E\right),
\label{eq:3bQC-1}
\end{eqnarray}
where the total energy of the dimer-particle system, $E$, is one of the discrete solutions of the QC
\begin{eqnarray}
\rm{Det}(1+\tilde{\mathcal{M}}^{\infty}_{V}\delta\tilde{\mathcal{G}}^V) = 0.
\label{eq:3bQC-2}
\end{eqnarray}
These energies are obtained from a LQCD computation of three-hadron correlation functions with the highest overlap with the s-wave channel. In Eq.~(\ref{eq:3bQC-1}), the incoming (outgoing) particle has momentum $\mathbf{p}$ $(\mathbf{k}$) and the incoming (outgoing) dimer has momentum $\mathbf{P}-\mathbf{p}$ $(\mathbf{P}-\mathbf{k})$, where $\mathbf{P}$ denotes the total momentum of the dimer-particle system. $\delta\tilde{\mathcal{G}}^V$ is a known kinematic FV function~\cite{Briceno:2012rv}. Diagrammatic expansions of the amplitudes $ \tilde{\mathcal{M}}^{\infty}_V$ and $ \tilde{\mathcal{M}}^{\infty}_{\infty}$ are depicted in the upper panel of Fig.~\ref{fig:3b}, and that of the full dimer propagator is depicted in the lower-right panel of the figure. $\delta{\mathcal{D}^{V}}$ in Eq.~(\ref{eq:3bQC-1}) is defined as $\delta\mathcal{D}^{V}=\mathcal{D}^{V}-\mathcal{D}^{\infty}$. The full s-wave dimer propagator in infinite volume can be written in terms of the physical scattering amplitude, and that in the finite volume includes a dependence on the FV function $c^{{q}^*}_{00}$, see Eq.~(\ref{eq:QC}). The complexity of this QC is due to the requirement of solving a sum/integral equation. Multitude of discrete energy eigenvalues are needed to constrain given parametrizations of the amplitude at low energies. 

Given the computational cost of multi-baryon calculations, it is crucial that LQCD computations be optimized to better constrain the output. One possibility is to examine the QC through a reverse practice prior to running the LQCD computation. This procedure extracts the expected spectra given an input dynamic and its associated uncertainties, thus provides the optimum parameters for the QCD-ensemble productions and correlator evaluation. The advantage of this reverse practice is highlighted in the two-nucleon sector as discussed in Sec.~\ref{sec:NN}. Ideally, one needs to find out how sensitive the energy spectra of three-nucleon systems are to the dynamics, and if more sensitivity can be gained by various strategies in light of inherently noisy calculations. These type of investigations have already been started in the meson sector~\cite{Mai:2018djl,Briceno:2018mlh}, and will continue to grow over the next few years.

%%%%%%%%%%%%%%%%%%%%%%%%%%%%%%%%%%%%%%%%%%%%%%%%%%%%%%%%%%%%%%%%%%%%%%%%%%%%%%%%%%%%%%%%%%%%%%%%%%%%%%
\section{Hadronic observables from background fields}
\begin{figure}[t!]
\begin{center}
\includegraphics[scale=0.525]{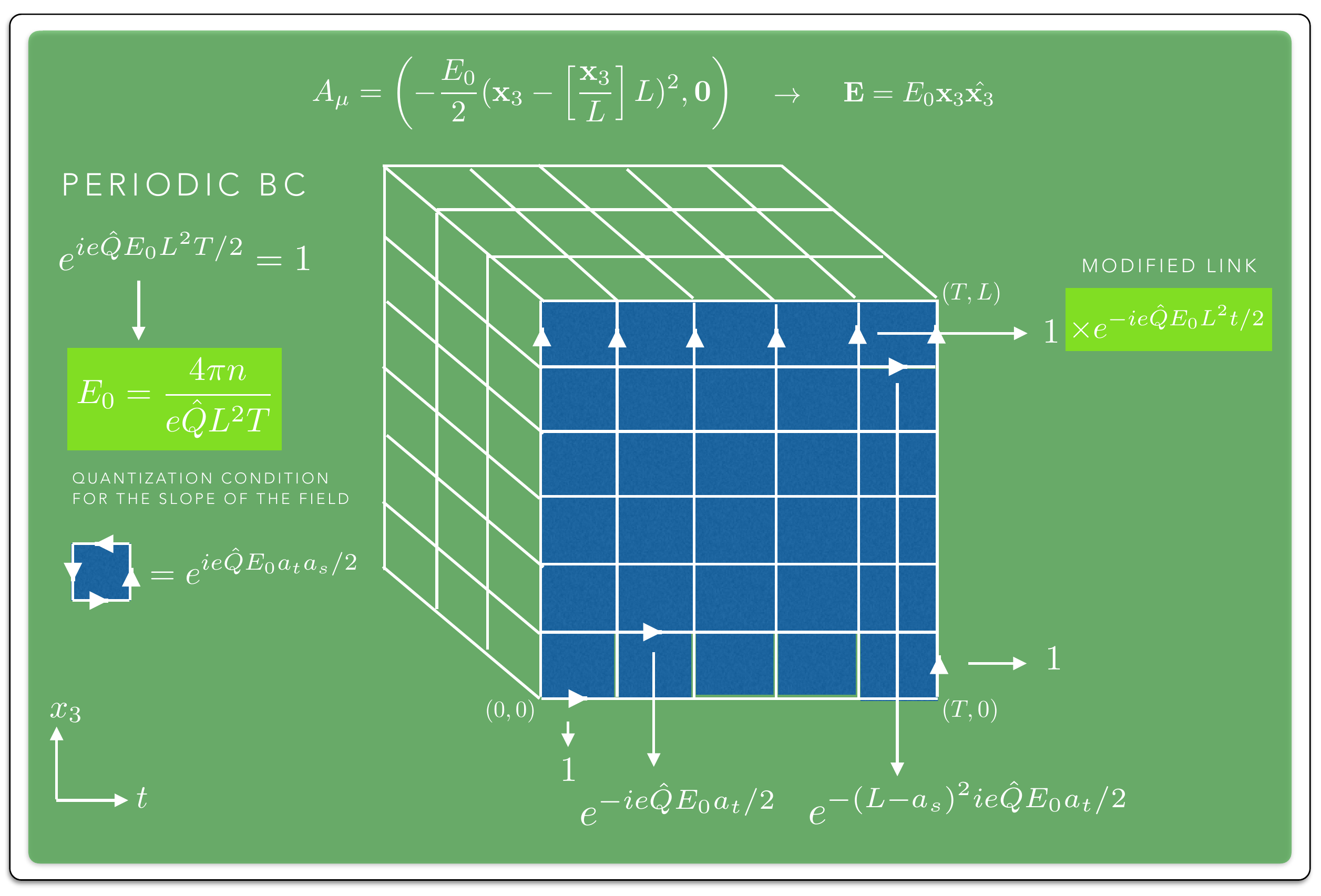}
\caption[.]{A 3D projection of a 4D lattice with periodic BCs. The values of the $U(1)$ gauge links for a fully periodic implementation of a linearly-rising electric field in the $\mathbf{x}_3$ direction are shown in the picture, along with the QC imposed on the slope of the electric field, $E_0$. $a_t$ ($T$) and $a_s$ ($L$) denote the lattice spacing (extent of the volume) along the temporal and spatial directions, respectively. $n$ is an integer and $e\hat{Q}$ is the charge of the quarks connected by the gauge link. See Ref.~\cite{Davoudi:2015cba} for further detail.}
\label{fig:bflinear}
\end{center}
\end{figure}
\noindent
Similar to experiment where certain EM structure properties of hadrons and nuclei can be probed using external EM fields, a LQCD calculation with a classical $U(1)$ gauge field can be set up such that the modification to the FV spectrum of the hadronic system can be related to its EM structure properties. Notable examples of the application of this method are the introduction of a uniform background  magnetic field to constrain magnetic moments~\cite{Bernard:1982yu, Martinelli:1982cb, Lee:2005ds, Aubin:2008qp, Detmold:2010ts, Beane:2014ora, Parreno:2016fwu} and uniform background electric and magnetic fields to constrain electric and magnetic polarizabilities~\cite{Fiebig:1988en, Christensen:2004ca, Lee:2005dq, Detmold:2006vu, Detmold:2009dx, Detmold:2010ts, Primer:2013pva, Lujan:2014kia, Beane:2014ora, Chang:2015qxa}. Further, background fields can provide access to transition amplitudes induced by either EM or weak currents, as was proposed in Ref.~\cite{Detmold:2004qn}, and successfully implemented to obtain the M1 transition rate in a radiative capture process in the two-nucleon system in Ref.~\cite{Beane:2015yha}, and in the $pp$ fusion process in Ref.~\cite{Savage:2016kon}. For isovector quantities, it suffices to introduce the background field such that only the valence quarks are affected, while isoscalar quantities require generating gauge-field configurations that are modified by the $U(1)$ gauge links.

More general background fields with given space and time variation can be of further value. Examples of these are linear electric fields to access the charge radius and the electric dipole moment~\cite{Davoudi:2015zda}, oscillatory background fields to access vacuum magnetic susceptibility~\cite{Bali:2015msa}, hadron form factors~\cite{Detmold:2004kw, Davoudi:2015cba}, and the doubly virtual forward Compton scattering amplitude~\cite{Agadjanov:2016cjc, Agadjanov:2018yxh}, and time-dependent fields to access spin polarizabilities~\cite{PhysRevD.47.3757, Detmold:2006vu, Lee:2011gz, Engelhardt:2011qq, Davoudi:2015cba}. For the case of a linearly-rising electric field, the FV correlation functions are formed in Ref.~\cite{Davoudi:2015zda} and are matched to infinite-volume correlation functions from an effective low-energy description of spin-0 and spin-1 hadrons/nuclei in presence of the varying field. This matching is complex due to both the nontrivial time dependence of correlation functions in an electric field and the spatial variation of the correlation functions along the direction of the field variation. These conditions, nonetheless, are available to constrain both the charge radius and electric dipole moment, of e.g., the deuteron, in future LQCD calculations with proposed background fields.
\begin{figure}[t!]
\begin{center}  
\includegraphics[scale=0.525]{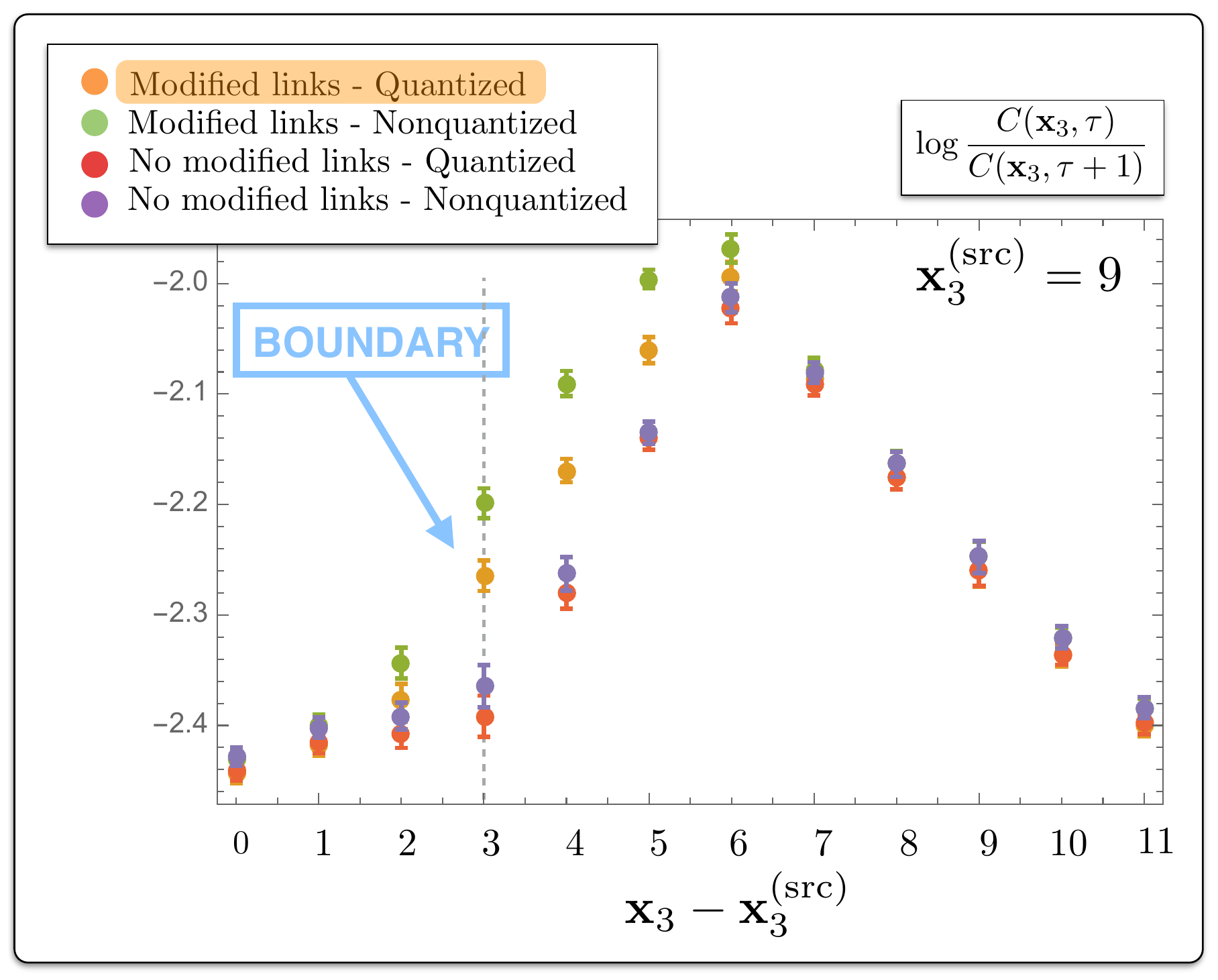}
\caption[.]{Dependence of the quantity $M(x_3,\tau) \equiv \log \frac{C(x_3,\tau)}{C(x_3,\tau+1)}$ on $x_3-x_3^{{\rm{(src)}}}$, formed out of the correlation function of the pion, $C(x_3,\tau)$ at a fixed time with various choices of background gauge fields that result in a linearly-rising electric field in the $\mathbf{x}_3$ direction. Only the implementation described in Fig.~\ref{fig:bflinear} gives rise to a smooth behavior at the boundary (denoted by the dashed line). All values are in units of lattice spacing. See Ref.~\cite{Davoudi:2015cba} for further detail.}
\label{fig:bfpbc}
\end{center}
\end{figure}

It is well known that in order to implement uniform background EM fields in a hypercubic lattice and maintain periodicity, certain modifications to the naive implementation of the gauge links on the lattice is required. Further, ``t'Hooft'' QCs must be imposed on the strength of the applied electric or magnetic field~\cite{'tHooft1979141, Smit:1986fn, AlHashimi:2008hr, Lee:2013lxa, Lee:2014iha}. A fully periodic implementation of background EM fields with sufficiently general space and time dependence is introduced in Ref.~\cite{Davoudi:2015cba}. As an example, consider the case of an electric field of the form $\mathbf{E}=E_0x_3 \mathbf{x}_3$, produced by the gauge potential $A_{\mu}=(-E_0x_3^2/2,0,0,0)$. As depicted in Fig.~\ref{fig:bflinear}, aside from the naive implementation in which the QCD gauge links are multiplied by the $U(1)$ gauge links of the form $e^{iA_0(x_3)a_t}$, one needs to impose two other requirements: i) the links along the $\mathbf{x}_3$ direction and adjacent to the boundary of lattice must be modified in the way shown, and ii) the slope of the electric-field strength must be quantized as stated in the figure. An exploratory LQCD calculations (with quenched gauge-field configurations and larger quark masses than in nature) verifies the periodicity of the procedure proposed. As is seen in Fig.~\ref{fig:bfpbc}, the (connected) correlation function of the neutral pion in presence of the electric field $\mathbf{E}=E_0x_3 \mathbf{x}_3$ exhibits nonuniformities near the boundary if only the requirement i) or ii), or neither, is implemented. Such nonuniformities introduce a systematic uncertainty into hadronic quantities that are hard to quantify. The fully periodic implementation of Fig.~\ref{fig:bflinear}, however, leads to a smooth behavior near the boundary, as is evident from Fig.~\ref{fig:bfpbc}.

%%%%%%%%%%%%%%%%%%%%%%%%%%%%%%%%%%%%%%%%%%%%%%%%%%%%%%%%%%%%%%%%%%%%%%%%%%%%%%%%%%%%%%%%%%%%%%%%%%%%%%
\section{Hadronic observables in presence of QED}
\begin{figure}[t!]
\begin{center}  
\includegraphics[scale=0.665]{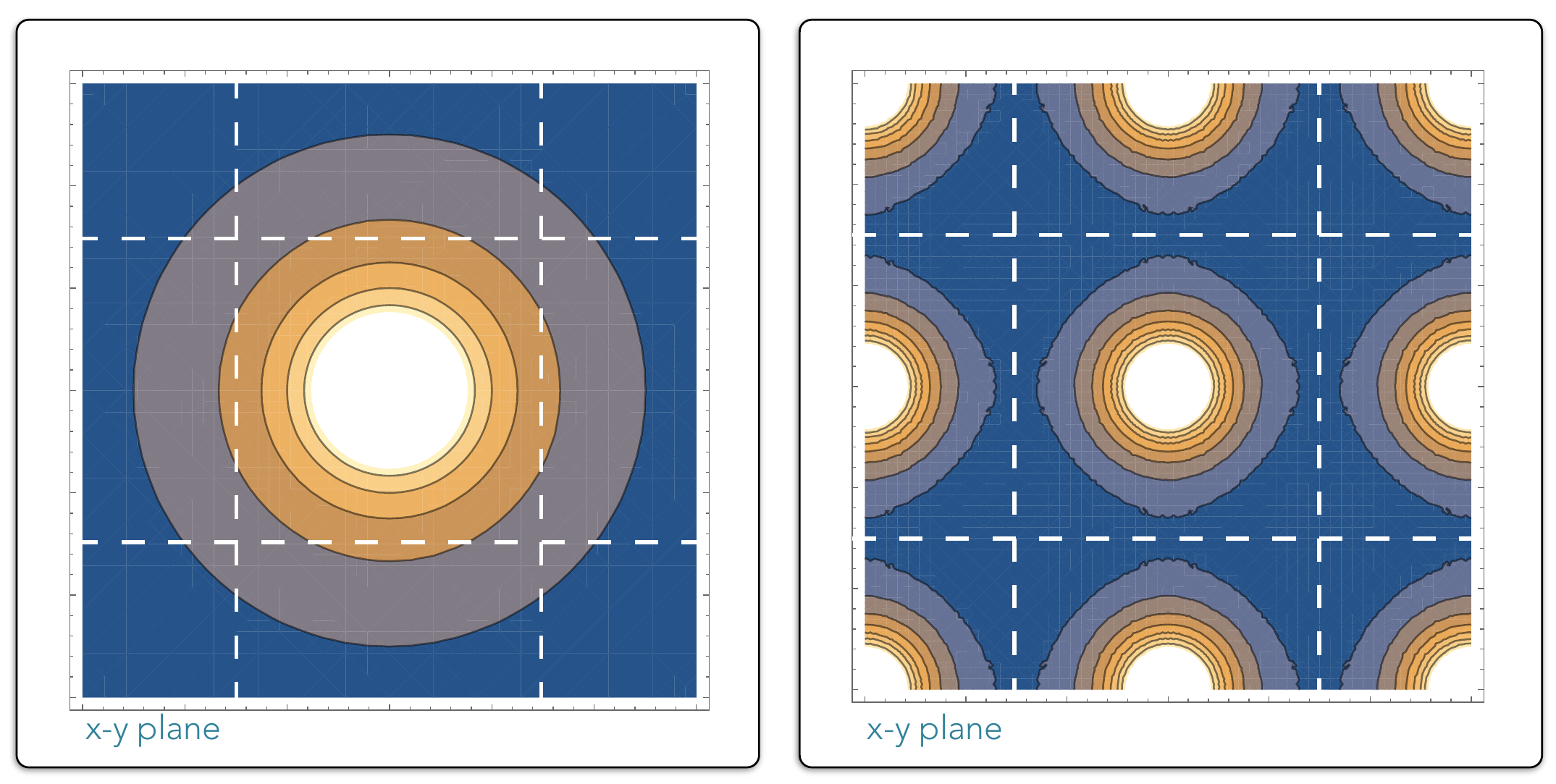}
\caption[.]{The contour plot of the Coulomb potential in the $x$-$y$ plane, produced by an electric charge at the center of the lattice volume, as is shown in the left panel, is inconsistent with periodic BCs. Removing the spatial zero mode of the photon, while it distorts the Coulomb potential significantly, is compatible with periodic BCs, as is shown in the right panel. The distortion in the potential can be analytically studied and corrected for, see Ref.~\cite{Davoudi:2014qua}. The periodic images of the volume are separated by the white dashed lines.}
\label{fig:potential}
\end{center}
\end{figure}
There has been a growing interest in augmenting LQCD studies with fully dynamical QED interactions. The primary drive is the need for more precise determinations of quantities in flavor physics for tests of the SM and searches for new physics. In particular, the uncertainty in state-of-the-art LQCD determinations of weak matrix elements in leptonic and semileptonic decay of mesons is now comparable to estimates of neglected isospin effects, including QED effects of $\mathcal{O}(\alpha) \sim 1\%$. For the nonleptonic decay of kaons, a $\sim10\%-20\%$ determination of the ratio of the direct ($\epsilon'$) and indirect ($\epsilon$) CP violation parameter with LQCD is in horizon. Given the nonperturbative enhancement of the isoscalar amplitude of pions over the isotensor one, the QED effects in ${\rm Re}(\epsilon'/\epsilon)$ may be much larger than the naive estimates. QED effects are also under investigation in LQCD studies of the muon anomalous magnetic moment~\cite{Boyle:2017gzv, Gulpers:2018mim, Blum:2017cer} and charged-hadron scattering. LQCD calculations of light nuclei towards the precision era will need to account for QED effects as well, as QED plays an important role in the binding of nuclei.

Given the zero mass of the photons, their dynamics can not be fully embedded in a finite volume of space. In particular, the Coulomb potential of a charged particle is incompatible with periodic BCs, as is shown in the left panel of Fig.~\ref{fig:potential}, and any attempt to impose periodic BCs will lead to the violation of Gauss's law~\cite{Hayakawa:2008an, Davoudi:2014qua}. One solution to this problem is to remove the spatial zero mode of the photon field, giving rise to a modified Coulomb potential,
\begin{eqnarray} 
V(\mathbf{r}-\mathbf{r}')=V^{(\infty)}(\mathbf{r}-\mathbf{r}')+\frac{1}{L^3}\sum_{\mathbf{m}\neq \mathbf{0}}\int\frac{d^3k}{(2\pi)^3}\frac{e^{i \mathbf{k} \cdot (\mathbf{r}-\mathbf{r}')}}{\mathbf{k}^2}e^{i\mathbf{k} \cdot \mathbf{m} L},
\label{eq:Vc}
\end{eqnarray}
which is plotted in the right panel of Fig.~\ref{fig:potential}, and exhibits full periodicity. $L$ in Eq.~(\ref{eq:Vc}) denotes the spatial extent of the cubic volume and $V^{(\infty)}$ is the infinite-volume potential. An interesting result is the corrections to the self energy of a classical charged sphere of radius $R$ and charge $e\hat{Q}$:
\begin{eqnarray} 
U(L) =  
{3\over 5} {(\hat{Q}e)^2\over 4 \pi R}+ 
{(\hat{Q}e)^2\over 8\pi L}\ c_1+ 
{(\hat{Q}e)^2\over 10 L} \left({R\over L}\right)^2 + \cdots,
\label{eq:Uc}
\end{eqnarray}
with the first term being the infinite-volume result. As was noted in Ref.~\cite{Davoudi:2014qua}, the LO volume correction is the same as that of the relativistic hadrons, and using a heavy-field EFT, similar structure-dependent terms at $\mathcal{O}(\alpha/L^3)$ is obtained for a hadron. These corrections have been recently evaluated in Ref.~\cite{Davoudi:2018qpl} for the self-energy of point-like charged particles on and off the mass shell and with a general boost vector $\mathbf{v}$. The first two orders are universal corrections, i.e., are independent of spin or structure of the hadron~\cite{Borsanyi:2014jba}, and can be written as:
\begin{eqnarray}\Delta\omega=e^2\hat{Q}m\left\{
\frac{1}{\gamma(\mathbf{v})}\frac{c_{2,1}(\mathbf{v})}{8\pi^2mL}+\frac{c_1}{4\pi m^2L^2}+\cdots\right\},
\label{eq:onshell}
\end{eqnarray}
where $m$ is hadron's mass, $\hat{Q}$ is its charge in units of $e$ and $\gamma(\mathbf{v})$ is the relativistic boost factor. $c_1=2.83729748\cdots$ and $c_{2,1}(\mathbf{v})$ is a boost-dependent regulated summation defined and evaluated in Ref.~\cite{Davoudi:2018qpl}.
 
So far, there has been tremendous progress in studies of mass splittings in hadron multiplets with the zero-mode removed formulation of QED in a FV, see Refs.~\cite{Portelli:2015wna, Patella:2017fgk} for recent reviews. Nonetheless, subtleties arising from such a nonlocal formulation may need to be studied more carefully when it comes to other observables and higher orders in an expansion in the fine structure constant. One such subtlety had already been discovered in the context of an EFT description of the hadrons coupled to zero-mode regulated QED, where a discrepancy in the mass shift in a finite volume was seen between the EFT and the full QED results at $\mathcal{O}\left(\alpha /L^3\right)$. While explanations, such as the need to introduce antiparticle degrees of freedom to the low-energy theory, are proposed~\cite{Fodor:2015pna, Lee:2015rua}, a more natural explanation consistent with the spirit of an EFT is provided recently in Ref.~\cite{Davoudi:2018qpl}. Since the removal of the zero mode is equivalent to introducing a uniform charge density over the entire volume, even though the high-energy degrees of freedom are integrated out in the EFT, fields still interact at short distances with the background charge. To capture the effect of these short-distance interactions, new operators must be introduced to the EFT, whose LECs can only be fit to a full QED calculation when possible, or a LQCD+LQED calculation when nonperturbative hadronic effects must be taken into account. Given that the charge density scales as $1/L^3$, the lowest-order operator in the Lagrangian coupled to this charge is necessarily a mass term, with a coefficient that can matched to the full QED calculation in the case of point-like particles:
\begin{eqnarray}
\delta m_{S=0}=-e^2\hat{Q}^2\frac{\mathbf{v}^2}{8m\omega(\mathbf{p}) L^3},~~~\delta m_{S=1/2}=e^2\hat{Q}^2\frac{2\omega(\mathbf{p})^2+m^2}{8m\omega(\mathbf{p})^3 L^3}.
\label{eq:resmass}
\end{eqnarray}
There are other approaches to implementing QED in a finite volume in a consistent manner that do not break locality, including the use of charged conjugation BCs and the massive photons. We refer the reader to recent literature on features of such implementations~\cite{Endres:2015gda, Bussone:2017xkb, Polley:1993bn, Wiese:1991ku, Kronfeld:1992ae, Kronfeld:1990qu, Lucini:2015hfa, Hansen:2018zre, Feng:2018qpx}.

Finally, it must be noted that developing the FV technology to extract physical observables from various formulations of QED in a finite volume is an active area of research. Studies of leptonic decay of mesons in presence of QED are in progress and FV features of these calculations, including the infrared divergence issue are understood~\cite{Carrasco:2015xwa, Lubicz:2016xro, Tantalo:2016vxk, Giusti:2018guw}. The nonrelativistic scattering of charged hadrons in a FV is studied in Ref.~\cite{Beane:2014qha} and a modified L\"uscher's formula is derived. Further, there are investigations into generalizing the Lellouch-L\"uscher formula to charged initial and final states, see e.g., Refs.~\cite{Christ:2017pze, Cai:2018why} and further developments will be expected in near future in this area.

%%%%%%%%%%%%%%%%%%%%%%%%%%%%%%%%%%%%%%%%%%%%%%%%%%%%%%%%%%%%%%%%%%%%%%%%%%%%%%%%%%%%%%%%%%%%%%%%%%%%%%
\section{Outlook}
\noindent
Exciting developments in LQCD research bring the prospect of making impactful contributions in advancing our theoretical knowledge of a diverse set of problems of phenomenological interest that have eluded predictions to date, continuing on a successful trend in delivering important results in hadronic and nuclear physics in recent years. Many of the interesting problems to be yet tackled are multi-particle in nature, introducing further layers of complexity as the number of particles increases, both in requiring significant computational resources and in formal mappings between finite and infinite-volume physics. Such increase in the number of particles is indeed inevitable towards the physical point, where pions are sufficiently light to be mass produced in processes, or when QED contributions are concerned in large volumes where multitude of photons can be radiated in a given process.

Formalisms that are put in place to address the mapping between FV spectra and physical amplitudes in the three-particle sector are becoming comprehensive and practical, however, the time scale of their development points to the complexity of the approaches involved, making a universal $N$-particle mapping a distant but much needed result. In the meantime, with the aim of relevant EFTs, QCD-based predictions in the many-body sector of hadronic and nuclear physics will not be out of reach. On one hand, direct mappings between observables in two, three and possibly four-hadron sector and LQCD output for energies and matrix elements are/may be a reality, and on the other hand, constrains on few-body observables can turn into constraints on the LECs of EFTs, which can be used in \emph{ab initio} calculations of many-body dynamics, a program that has already started~\cite{Barnea:2013uqa, Kirscher:2015yda, Contessi:2017rww, Kirscher:2017fqc, Bansal:2017pwn}.

It will be interesting to see how this field will evolve and whether breakthroughs, such as that introduced by Martin L\"uscher over three decades ago, will come along to change the course of the field towards maturity and versatility. With such open questions, it is an exciting time to be in this field, and to work towards filling in the gaps between significant numerical results that computational and algorithmic advancements have enabled us to reach, and how far our formal and conceptual understanding permits us to go to interpret these results.

%%%%%%%%%%%%%%%%%%%%%%%%%%%%%%%%%%%%%%%%%%%%%%%%%%%%%%%%%%%%%%%%%%%%%%%%%%%%%%%%%%%%%%%%%%%%%%%%%%%%%%
\section{Acknowledgment}
\noindent
I would like to thank the 2018 Wilson Award Committee, the International Advisory Committee and the Local Organizing Committee of the 36th Annual International Symposium on Lattice Field Theory at Michigan State University, for giving me the opportunity to present on advancements in finite-volume technologies for lattice field theories; advancements that are the results of a coherent, vigorous and directed effort by a number of lattice gauge theorists and effective field theorists over the years. I am, in particular, indebted to all my mentors and/or co-authors over the years, including Martin Savage, Silas Beane, Ra\`ul Brice\~no, Emmanuel Chang, William Detmold, Michael Endres, James Harrison, Marc Illa, Andreas J\"uttner, David Kaplan, Thomas Luu, Kostas Orginos, Assumpta Parre\~no, Antonin Portelli, Neda Sadooghi, Steve Sharpe, Phiala Shanahan, Brian Tiburzi and Michael Wagman, for scientific inspiration and wonderful collaborations.

\bibliographystyle{JHEP}
\bibliography{bibi}

\end{document}